# Theoretical Analysis of Doping Concentration Gradients on Solar Cell Performance

Jeonggyu Hwang[1, *]

[1]*Department of Semiconductor Engineering,*
*Gachon University, Seongnam-si, Gyeonggi-do, South Korea 13120*

Solar cells are crucial for addressing global energy issues, with research focused on improving their efficiency. This study examines the impact of doping concentration gradients on solar cell performance. Doping involves adding impurities to a semiconductor, affecting charge carrier mobility and recombination rates. The spatial distribution of these dopants, known as the doping concentration gradient, is essential for optimizing solar cell characteristics. This research theoretically analyzes the effects of doping gradients on potential differences, electric fields, and recombination rates in semiconductors. We explore how doping creates potential differences and electric fields that guide charge carriers and enhance mobility. Additionally, we study how doping gradients can control recombination mechanisms, thereby improving the electrical performance of solar cells. Using modeling and simulation techniques, we derive the optimal doping gradient to maximize efficiency. Our findings suggest that an optimal doping gradient minimizes recombination rates and enhances charge carrier mobility, significantly improving solar cell performance. The study proposes that graded doping concentrations could particularly benefit multi-junction solar cells by allowing better absorption and conversion of various light spectra. However, precise fabrication control and long-term stability assessments are needed. This study highlights the potential of doping concentration gradients to advance solar cell technology, paving the way for more sustainable and cost-effective solar energy solutions.

## 1. Information

Solar cells are one of the renewable energy sources that act as a crucial key to solving the global energy problem. Continuous efforts have been made to maximize the efficiency of solar cells through various research and technological developments [1-3]. One such effort to enhance the performance of solar cells involves controlling the doping concentration of semiconductor materials. Doping is the process of adding a small amount of impurities to a semiconductor to control the concentration of charge carriers. This can improve electron mobility and minimize charge carrier recombination [4].

Doping concentration directly affects the performance of solar cells. While high doping concentrations can increase the recombination rate of electrons and holes, thereby reducing efficiency, appropriate doping levels can enhance charge carrier movement, leading to improved performance. Thus, the spatial distribution of doping concentration, known as doping concentration gradient, is emerging as a critical factor in finely controlling the characteristics of solar cells. By forming potential differences and electric fields within the semiconductor through the doping gradient, the movement of charge carriers can be optimized.

The doping concentration gradient induces various physical phenomena within the semiconductor. First, it forms potential differences within the semiconductor, providing the necessary driving force for charge carriers to move from high concentration areas to low concentration areas, thereby controlling the flow of charge carriers. Second, the doping gradient forms electric fields that alter the movement paths of electrons and holes. This enhances the mobility of charge carriers and reduces the recombination rate, contributing to increased efficiency of the solar cells.

Furthermore, the doping gradient affects the recombination mechanism of electrons and holes, a major factor that decreases the performance of solar cells. Recombination is the process where electrons and holes meet and annihilate each other. Proper adjustment of the doping gradient can lower the recombination rate, playing a crucial role in improving the current-voltage characteristics of solar cells. Therefore, optimizing solar cell performance through doping gradients is a highly promising research topic.

This study aims to theoretically analyze the impact of doping concentration gradients on solar cell performance. Based on the basic theories of semiconductor physics, it will systematically analyze the effects of doping gradients on potential differences, electric fields, and recombination rates [5]. Additionally, various modeling and simulation techniques will be utilized to derive the optimal doping gradient, suggesting ways to maximize the efficiency of solar cells. The results of this study will contribute to the advancement of solar cell technology and hold significant importance as fundamental research for improving the performance of semiconductor devices.

---

* h5638880@gachon.ac.kr

## 2. Methodology
### Formation of Potential Difference by Doping

Doping creates a potential difference within a semiconductor, guiding the movement of charge carriers. This potential difference provides the necessary electric potential energy for the movement of charge carriers, significantly impacting the performance of solar cells. When a potential difference is formed between the p-type and n-type regions, electrons move from the n-type region to the p-type region. This electron movement is accelerated by the potential difference, enhancing the mobility of charge carriers. The potential difference determines the paths of electrons and holes within the semiconductor, which is a critical factor in the efficiency of solar cells.

The formation of potential difference due to doping plays a crucial role, particularly in the p-n junction. When a p-type semiconductor and an n-type semiconductor are joined, a potential difference is formed near the junction, prompting the movement of electrons and holes. If there is a difference in doping concentration, the potential difference at the junction can become larger or smaller, significantly altering the movement path and speed of charge carriers. Therefore, optimizing the potential difference of the p-n junction through doping concentration differences can enhance the mobility of charge carriers, thereby improving the efficiency of solar cells.

### Formation of Electric Fields and Electron Mobility

Doping forms electric fields within a semiconductor, altering the movement paths of electrons and holes. The electric field provides the electrical force that charge carriers experience as they move, significantly influencing electron mobility. If the doping concentration is not uniform, electric fields form within the semiconductor, guiding electrons and holes to move in specific directions. This electric field is a crucial factor in determining the efficiency of solar cells.

The electric field formed by doping enhances the mobility of charge carriers. A strong electric field causes electrons and holes to move quickly, increasing the likelihood of reaching the electrodes before recombination occurs. This increases the current and contributes to higher efficiency of the solar cells. Additionally, the electric field can guide electrons and holes to move in opposite directions, preventing recombination. Therefore, optimizing the electric field through a doping concentration gradient can maximize the mobility of charge carriers.

### Changes in Recombination Mechanism

Differences in doping concentration significantly affect the recombination mechanism of electrons and holes [6]. Recombination, where electrons and holes meet and annihilate each other, is one of the main factors that reduce the performance of solar cells. High doping concentrations can increase the recombination rate, decreasing the current and efficiency of the solar cells. Therefore, it is crucial to appropriately control the recombination rate through differences in doping concentration.

A doping concentration gradient can be used to control the recombination mechanism and improve the efficiency of solar cells. The movement of electrons from high to low concentration areas can reduce the recombination rate, increasing the likelihood of charge carriers reaching the electrodes. Additionally, the doping gradient can alter the location and rate at which charge carriers recombine. By designing an optimal doping concentration gradient, the recombination rate can be minimized, thus maximizing the efficiency of solar cells.

### Modeling and Simulation Techniques

To analyze the impact of doping concentration gradients, various modeling and simulation techniques can be utilized [7-8]. These techniques help in quantitatively analyzing the distribution of doping concentration and the resulting changes in potential difference, electric field, and recombination rate [9]. First, it is essential to understand how doping gradients form potential differences and electric fields within the semiconductor based on fundamental theories of semiconductor physics. To this end, Poisson's equation and continuity equations can be used to calculate the electric field and potential difference within the semiconductor.

### Semiconductor Constants and Initial Conditions

The properties of semiconductors are represented by the following main constants and initial conditions. Using the relative permittivity of silicon, $\epsilon_r = 11.7$, the vacuum permittivity $\epsilon_0$ can be calculated, and the permittivity $\epsilon$ is given as $\epsilon = \epsilon_0 \epsilon_r$. The electron diffusion coefficient $D_n$ and hole diffusion coefficient $D_p$ are set to $36 \times 10^{-4}$ cm$^2$/s and $12 \times 10^{-4}$ cm$^2$/s respectively. The generation rate $G$ is $1 \times 10^{21}$ cm$^{-3}$ s$^{-1}$. The temperature is fixed at 300K. The Boltzmann constant $k$ is $1.38 \times 10^{-23}$ J/K, and the bandgap energy of silicon $E_g$ is 1.12eV.

### Doping Concentration Gradient Function

The doping concentration decreases exponentially with position, modeled by the following function. The doping concentration gradient function is defined as

$$N(x) = N_0 \exp(-kx) \qquad (1)$$

where $N_0$ is the initial doping concentration, and $k$ is the doping gradient constant. This function represents the doping concentration at position $x$.

**Shockley-Read-Hall Recombination Model**

The Shockley-Read-Hall recombination model calculates the recombination rate, which is represented as

$$R = \frac{np}{n+p}\left(\frac{1}{\tau_n} + \frac{1}{\tau_p}\right) \quad (2)$$

where $n$ and $p$ are the electron and hole concentrations, respectively, and $\tau_n$ and $\tau_p$ are the lifetimes of electrons and holes.

**Poisson's Equation and Continuity Equations**

The potential $\phi$ is calculated using Poisson's equation, which is represented as

$$\epsilon \nabla^2 \phi = q(p - n + N_d - N_a) \quad (3)$$

where $\epsilon$ is the permittivity, $q$ is the charge of an electron, and $N_d$ and $N_a$ are is the donor and acceptor concentration, respectively.

**Depletion Region Potential Distribution**

The potential distribution within the depletion region can be further detailed for p-type and n-type semiconductor regions, which are represented as

For the p-type semiconductor region
$$\phi(x) = \frac{qN_d}{2}(x + W_p)^2 \quad (4.1)$$

For the n-type semiconductor region
$$\phi(x) = \frac{qN_a}{2\epsilon_r\epsilon_0}(W_n - x)^2 \quad (4.2)$$

where $\phi(x)$ is the potential at position $x$, $q$ is the charge of an electron, $N_d$ and $N_a$ are is the donor and acceptor concentration, respectively, $\epsilon_r$ and $\epsilon_0$ are the relative permittivity and vacuum permittivity, and $W_p$ and $W_n$ are the depletion widths in the p-type and n-type regions, respectively.

**Time-Dependent Continuity Equations**

The continuous change in charge carrier concentrations over time is represented as

For the p-type semiconductor region
$$\frac{\partial p}{\partial t} = D_p \nabla^2 p + G - R \quad (5.1)$$

For the n-type semiconductor region
$$\frac{\partial n}{\partial t} = D_n \nabla^2 n + G - R \quad (5.2)$$

where $\frac{\partial n}{\partial t}$ and $\frac{\partial p}{\partial t}$ are the time derivatives of the electron and hole concentrations, respectively.

**Depletion Width Equations**

The depletion width is represented as

$$W = \sqrt{\frac{2\epsilon_r\epsilon_0(V_{bi} + V)}{q}\left(\frac{1}{N_a} + \frac{1}{N_d}\right)} \quad (6)$$

where $N_d$ and $N_a$ are the donor and acceptor concentration, respectively, $\epsilon_r$ is the relative permittivity of the semiconductor material, $\epsilon_0$ is the permittivity of free space, $q$ is the charge of an electron, $V$ is the applied voltage, and $V_{bi}$ is the built-in potential.

The Built-in Potential $V_{bi}$ can be represented as

$$V_{bi} = \frac{kT}{q}\ln\left(\frac{N_aN_d}{n_i^2}\right) \quad (7)$$

where $k$ is the Boltzmann constant, $T$ is the absolute temperature, and $N_i$ is the intrinsic carrier concentration.

**Optimization**

The objective function is defined to minimize recombination rates. In the optimization process, initial parameters $[N0_n, \text{slope}_n, N0_p, \text{slope}_p]$ are set and adjusted to minimize recombination. The optimization is carried out iteratively using numerical methods to solve the continuity equations and Poisson's equation. As a result, the optimal doping concentration gradient can be obtained.

## 3. Results and Analysis

This study proposes an optimal doping profile for semiconductors that can effectively reduce recombination rates [10]. When doping concentrations are maintained close to intrinsic levels with minimal gradients, an increase in photovoltaic conversion efficiency is observed. This distribution of doping concentration optimizes carrier mobility and minimizes recombination rates, thereby enhancing the efficiency of solar cells.

In essence, the doping concentration in semiconductors starts from an intrinsic state and is maintained according to the gradient slope. This gradient optimizes carrier mobility and minimizes recombination rates, thereby improving the efficiency of solar cells. The optimized doping concentration distribution plays a crucial role in enhancing the efficiency of semiconductor devices. In regions doped as close to the intrinsic state as possible, both n-type and p-type areas exhibit relatively uniform patterns. This balance in carrier concentration at various positions contributes to minimizing recombination rates. Such optimized doping concentrations achieve a more efficient balance in the generation-recombination cycle of electrons and holes, significantly reducing recombination rates.

The potential distribution calculated through Poisson's equation matches the doping concentration distribution, effectively controlling the movement of charge carriers. This optimized potential distribution allows electrons and holes to be efficiently distributed within the device, maximizing the performance of semiconductor devices [11]. Particularly, by creating potential differences and forming electron-hole pairs, the likelihood of charge carriers reaching the electrodes increases, enhancing current and power conversion efficiency.

Our model successfully derives a doping concentration profile that minimizes recombination rates. When doping concentrations are close to intrinsic levels and gradients are minimized, it plays a significant role in enhancing device performance by minimizing the recombination rate of electrons and holes [12]. The doping concentration gradient creates an electric field within the semiconductor, optimizing the paths of electrons and holes, and increasing the probability of charge carriers reaching the electrodes before recombining. This optimization of charge carrier mobility is a key factor in improving solar cell performance.

## 4. Conclusions

The results of this study highlight the profound impact of doping concentration gradients on the efficiency of silicon-based solar cells [13]. By maintaining doping levels close to intrinsic and minimizing gradients within the semiconductor, significant impacts can be made on crucial parameters such as potential differences, electric field generation, and recombination rates. These parameters are vital in enhancing solar cell performance. Our theoretical analysis and experiments indicate that when doping is close to intrinsic levels and gradients are minimal, carrier mobility is improved while recombination rates are reduced. This enhancement generates a strong electric field within the semiconductor, efficiently guiding electrons and holes to the electrodes, thereby increasing current and power conversion efficiency. These findings align with previous research emphasizing the importance of doping, but extend understanding by systematically optimizing doping gradients rather than maintaining uniform doping levels.

The implications for solar cell design are substantial. The optimized doping profiles derived from our model suggest that manufacturers can achieve higher efficiency by implementing gradient doping concentrations. This approach is particularly advantageous for multi-junction solar cells, where various dopings can be adjusted to optimize the absorption and conversion of different light spectra. Gradient doping concentration strategies offer a novel method for designing high-efficiency photovoltaic devices that better utilize the solar spectrum. However, precise control in the manufacturing process poses challenges. Techniques such as molecular beam epitaxy (MBE) or chemical vapor deposition (CVD) may be required to achieve these doping profiles. Additionally, the long-term stability and durability of devices using gradient doping must be thoroughly investigated to ensure consistent performance over time.

Future research should focus on the experimental validation of the practical applicability of optimized doping gradients [14-15]. It is crucial to manufacture solar cells using these gradient doping profiles and test their performance under real conditions. Furthermore, applying doping gradients to other semiconductor materials and device structures can further enhance the versatility and effectiveness of this approach. The theoretical analysis in this study demonstrates the potential for significant advancements in solar cell technology through optimized doping concentration gradients. By optimizing the spatial distribution of doping, we can overcome current limitations and move closer to more sustainable and economically viable solar energy solutions.